\documentclass[conference]{IEEEtran}
\IEEEoverridecommandlockouts
\usepackage{cite}
\usepackage{amsmath,amssymb,amsfonts}
\usepackage{algorithmic}
\usepackage{graphicx}
\usepackage[capitalise,nameinlink,noabbrev]{cleveref}
\usepackage{textcomp}
\usepackage{booktabs}
\usepackage{url}
\usepackage{multicol}
\usepackage{xcolor}
\def\BibTeX{{\rm B\kern-.05em{\sc i\kern-.025em b}\kern-.08em
    T\kern-.1667em\lower.7ex\hbox{E}\kern-.125emX}}

\usepackage{glossaries}

\newcommand\todo[1]{\textcolor{black}{#1}\PackageWarning{}{#1!}}
\newcommand\changes[1]{\textcolor{black}{#1}\PackageWarning{}{#1!}}

\newif\ifreviewmode

\newacronym{asic}{ASIC}{application-specific integrated circuit}
\newacronym{dft}{DFT}{design for testability}
\newacronym{dz}{DZ}{Microelectronics Design Center at ETH Zurich}
\newacronym{fpga}{FPGA}{field-programmable gate array}
\newacronym{ic}{IC}{integrated circuit}
\newacronym{iis}{IIS}{Integrated Systems Laboratory}
\newacronym{snr}{SNR}{signal-to-noise ratio}
\newacronym{hpc}{HPC}{High Performance Computing}
\newacronym{ppa}{PPA}{Power Performance Area}
\newacronym{cmos}{CMOS}{complementary metal-oxide semiconductor}
\newacronym{simd}{SIMD}{Single Instruction Multiple Data}
\newacronym{sse}{SSE}{Streaming SIMD Extensions}
\newacronym{avx}{AVX}{Advanced Vector Extensions}
\newacronym{sve}{SVE}{Scalable Vector Extension}
\newacronym{rvv}{RVV}{RISC-V Vector Extension}
\newacronym{fpu}{FPU}{Floating Point Unit}
\newacronym{vrf}{VRF}{Vector Register File}
\newacronym{rtl}{RTL}{Register Transfer Level}
\newacronym{xbar}{XBAR}{Crossbar}
\newacronym{cmg}{CMG}{Core Memory Group}
\newacronym{vpp}{VPP}{Vector Parallel Pipeline}
\newacronym{alu}{ALU}{Arithmetic Logic Unit}
\newacronym{fma}{FMA}{Floating Multiply Accumulate}
\newacronym{llc}{LLC}{Last Level Cache}
\newacronym{hbm}{HBM}{High Bandwidth Memory}
\newacronym{vpu}{VPU}{Vector Processing Unit}
\newacronym{vlen}{VLEN}{Vector Length}
\newacronym{dlen}{DLEN}{Datapath Width}
\newacronym{elen}{ELEN}{Element Length}
\newacronym{isa}{ISA}{Instruction Set Architecture}
\newacronym{csr}{CSR}{Control Status Register}
\newacronym{vl}{VL}{Vector Length}
\newacronym{ew}{EW}{Element Width}
\newacronym{sldu}{SLDU}{Slide Unit}
\newacronym{tlp}{TLP}{Thread-Level Parallelism}
\newacronym{masku}{MASKU}{Mask Unit}
\newacronym{vlsu}{VLSU}{Vector Load Store Unit}
\newacronym{vldu}{VLDU}{Vector Load Unit}
\newacronym{vstu}{VSTU}{Vector Store Unit}
\newacronym{axi}{AXI}{Advanced eXtensible Interface}
\newacronym{addrgen}{ADDRGEN}{Address Generator}
\newacronym{sram}{SRAM}{Static Random Access Memory}
\newacronym{lmul}{LMUL}{Length Multiplier}
\newacronym{tt}{TT}{Typical Typical}
\newacronym{ss}{SS}{Slow Slow}
\newacronym{vcd}{VCD}{Value Change Dump}
\newacronym{ml}{ML}{Machine Learning}
\newacronym{ai}{AI}{Artificial Intelligence}
\newacronym{riscv}{RISC-V}{Reduced instruction set Computer}
\newacronym{DP}{DP}{double-precision}
\newacronym{dnn}{DNN}{Deep Neural Networks}
\newacronym{PPA}{PPA}{power, performance, and area}
\newacronym{GLSU}{GLSU}{Global Load Store unit}
\newacronym{RINGI}{RINGI}{Ring Interface}
\newacronym{REQI}{REQI}{Request Interface}
\newacronym{A2A}{A2A}{all-to-all}
\newacronym{dlp}{DLP}{Data-Level Parallelism}
\newacronym{VRF}{VRF}{Vector Register File}
\newacronym{MM}{MM}{matrix-matrix multiplications}
\newacronym{MV}{MV}{matrix-vector multiplications}
\newacronym{DOTP}{DOTP}{dot-product}
\newacronym{AXPY}{AXPY}{A.X+Y}
\newacronym{llm}{LLM}{Large Language Model}
\newacronym{gpu}{GPU}{Graphics Processing Unit}
\newacronym{gemm}{GEMM}{General Matrix Multiplication}
\newacronym{gemv}{GEMV}{General Matrix-Vector Multiplication}
\newacronym{blas}{BLAS}{Basic Linear Algorithm Program}
\newacronym{vpe}{VPE}{Vector Processing Element}
\newacronym{cpu}{CPU}{Central Processing Unit}
\newacronym{cc}{CC}{Core Complex}
\newacronym{fpr}{FPR}{Floating-Point Register File}
\newacronym{vfu}{VFU}{Vector Functional Unit}
\newacronym{tcdm}{TCDM}{Tightly Coupled Data Memory}
\newacronym{sta}{STA}{Static Timing Analysis}

\begin{document}

\title{TROOP: At-the-Roofline Performance for Vector Processors on Low Operational Intensity Workloads
\ifreviewmode
\else
\thanks{\textcopyright 2025 IEEE. Personal use of this material is permitted. Permission from IEEE must be
obtained for all other uses, in any current or future media, including
reprinting/republishing this material for advertising or promotional purposes, creating new
collective works, for resale or redistribution to servers or lists, or reuse of any copyrighted
component of this work in other works.}
\fi
}

\ifreviewmode
\author{\emph{Hidden for double-blind review purposes.}}
\else
\author{\IEEEauthorblockN{Navaneeth Kunhi Purayil\textsuperscript{1}, Diyou Shen\textsuperscript{1}, Matteo Perotti\textsuperscript{1}, and Luca Benini\textsuperscript{1,2}}
\IEEEauthorblockA{\textsuperscript{1}\textit{ETH Zürich, Zürich, Switzerland}, \textsuperscript{2}\textit{Università di Bologna, Bologna, Italy}}

\{nkunhi, dishen, mperotti, lbenini\}@iis.ee.ethz.ch
}
\fi

\maketitle

\bstctlcite{IEEE:BSTcontrol}

\begin{abstract}
The fast evolution of \gls{ml} models requires flexible and efficient hardware solutions as hardwired accelerators face rapid obsolescence.
Vector processors are fully programmable and achieve high energy efficiencies by exploiting data parallelism, amortizing instruction fetch and decoding costs.
Hence, a promising design choice is to build accelerators based on shared L1-memory clusters of streamlined \glspl{vpe}.
However, current state-of-the-art VPEs are limited in L1 memory bandwidth and achieve high efficiency only for computational kernels with high data reuse in the \gls{vrf}, such as \gls{gemm}. 
Performance is sub-optimal for workloads with lower data reuse like \gls{gemv}.
To fully exploit available bandwidth at the L1 memory interface, the \gls{vpe} micro-architecture must be optimized to achieve near-ideal utilization, i.e., to be as close as possible to the L1 memory roofline (at-the-roofline).
In this work, we propose TROOP, a set of hardware optimizations that include decoupled load-store interfaces, improved vector chaining, shadow buffers to hide VRF conflicts, and address scrambling techniques to achieve at-the-roofline performance for VPEs without compromising their area and energy efficiency.
We implement TROOP on an open-source streamlined vector processor in a 12nm FinFET technology. 
TROOP achieves significant speedups of 1.5$\times$, 2.2$\times$, and 2.6$\times$, respectively, for key memory-intensive kernels such as GEMV, DOTP and AXPY, achieving at-the-roofline performance.
Additionally, TROOP enhances the energy efficiency by up to 45\%, reaching 38 DP-GFLOPs/W (1 GHz, TT, 0.8V) for DOTP while maintaining a high energy efficiency of 61 DP-GFLOPs/W for \glspl{gemm}, incurring only a minor area overhead of less than 7\%.

\end{abstract}

\glsresetall

\begin{IEEEkeywords}
Vector processors, RISC-V, Efficiency
\end{IEEEkeywords}

\section{Introduction}
\label{sec:intro}

Recent advances in \gls{ai} have led to a significant increase in computation and memory requirements \cite{ai_memorywall}.
As single-core performance improvements through technology scaling approaches the physical and thermal limits, massively parallel architectures and domain-specific acceleration have been the main approaches to achieve higher performance and efficiencies. 
These approaches are particularly effective with state-of-the-art workloads such as \glspl{llm}, which exhibit high degrees of data parallelism.

Today, the attention layers of \glspl{llm} are the most dominant workload, contributing to about 70-80\% of operations during inference \cite{yuan2025native}.
In the initial prefill phase, the attention layer consists of \gls{gemm} operations with high operational intensity, motivating the design of accelerators optimized for low bandwidth-to-compute ratios.
Various accelerators have been designed to accelerate compute-intensive machine learning workloads, primarily matrix multiplications and convolutions \cite{speed_dnn, eyerissv2, flexnn}. 
In contrast, the decode phase that follows the prefill phase consists of \gls{gemv} operations, which have low operational intensity, mapping accesses to a query-specific Key-Value cache with limited data reuse. Thus, architectures require higher memory bandwidth to accelerate this operation.
Although these low-level operations are common across many machine learning models, the rapid evolution of AI algorithms demands flexible computing architectures that can adapt to emerging workloads while maintaining high performance and energy efficiency.

Vector processor architectures have seen a resurgence in recent years, driven by their high performance, energy efficiency, and well-known programming model.
By encoding multiple operations in a single instruction, they amortize the instruction fetch, decode, and issue energy cost. Additionally, the vector-\gls{simd} programming model enables easy software development and porting across various hardware implementations.
Hence, a common design pattern consists of a many-core shared L1-memory cluster, with each core being an efficient, streamlined \gls{vpe}, offering an advantageous compromise between flexibility and efficiency, and providing a clear scale-out path with hierarchical architectures featuring multiple many-core clusters.

However, today's state-of-the-art \glspl{vpe} suffer from insufficient L1-memory bandwidth to achieve peak \gls{fpu} utilizations when executing low operational intensity workloads. 
For example, while current \glspl{vpe} typically offer a maximum bandwidth-to-compute ratio of 1:1, memory-intensive kernels such as DOTP and AXPY require ratios of 2:1 and 3:1, respectively, to fully utilize the \gls{fpu}.

Shared L1 memory clusters typically offer greater bandwidth in L1 memory than what is currently utilized by today’s VPEs, a gap that will widen with advances in memory technologies, particularly the shift toward full 3D integration \cite{tsmc3d, cmos2.0}.
Consequently, designing VPEs that can fully exploit this growing bandwidth can significantly enhance the performance of low operational intensity workloads, and, by extension, improve the overall efficiency of applications that include both memory- and compute-intensive phases.

In this paper, we propose TROOP, a comprehensive set of micro-architectural optimizations designed to increase and fully utilize the memory bandwidth of \glspl{vpe}, achieving near-peak performance for low operational intensity workloads. We evaluate TROOP on the open-source, RVV-based Spatz \glspl{vpe}, demonstrating that simply increasing memory bandwidth by widening the load/store ports into L1 is insufficient to achieve peak \gls{fpu} utilization on memory-intensive kernels, due to the resulting imbalance between memory bandwidth and compute throughput. To overcome this bottleneck, TROOP introduces a range of enhancements--including improved dependency handling, instruction chaining, and optimized register file and memory access patterns--that collectively enable performance near the roofline, i.e., approaching the theoretical L1 memory bandwidth limit, for memory-intensive workloads.

The main contributions of this work are:
\begin{itemize}
    \item Design of TROOP, a set of micro-architectural optimizations, including decoupled \gls{vlsu}, improved vector chaining, shadow buffers to hide \gls{vrf} conflicts, and address scrambling techniques to increase and fully utilize the bandwidth of \glspl{vpe} and achieve performance at the L1-memory roofline for memory-intensive kernels.
    \item Integration and evaluation of TROOP on the open-source Spatz \gls{vpe} showing significant speedups of up to 2.6$\times$ (AXPY), reaching at-the-roofline performance for both high and low operational intensity kernels.
    \item Evaluation of TROOP's impact on VPEs' PPA metrics. In 12nm FinFET technology, TROOP increases Spatz's energy efficiency by up to 45\% when executing a memory-intensive DOTP kernel, achieving 38 DP-GFLOPs/W (1 GHz, TT, 0.8V), with a small area overhead of less than 7\%. 
\end{itemize}

\section{Related works}
\label{sec:relworks}

\gls{ml} accelerators traditionally employ extremely specialized architectures to maximize applications' performance and energy efficiency, often focusing on the ubiquitous matrix multiplication and convolution kernels \cite{speed_dnn, eyerissv2, flexnn}.
However, the rapidly changing \gls{ai} landscape has also driven the industry focus towards more flexible solutions.
This trend is testified by cloud inference platforms \cite{allinone, tensix, mi300x} featuring scalar, vector, and matrix units within a core.
While the scalar core executes the bookkeeping code and the matrix accelerator computes the operationally intensive matrix multiplications, the vector unit handles the usually more memory-intensive vector kernels.

However, conventional vector processors are often designed with low bandwidth-to-compute ratios targeting peak \gls{fpu} utilizations on computationally intensive kernels. 
For example, Spatz, Saturn, and Vitruvius+ \cite{cavalcante2023spatz, saturn, minervini_vitruvius_2023} feature a bandwidth-to-compute ratio of 1:1, while Ara \cite{ara2perotti24} and the vector core of NEC VE30 \cite{Takahashi_2023} have lower ratios of 1:2 and 1:6, respectively. Although these vector processors can achieve near-ideal \gls{fpu} utilizations on compute-intensive kernels, they cannot fully exploit their \glspl{fpu} on kernels with lower operational intensity (e.g., DOTP, AXPY, GEMV) due to their insufficient memory bandwidth, resulting in low overall performance. 

Research efforts have been made to overcome the memory wall, improve the performance and efficiency of \gls{vpe} architectures.
In the traditional von Neumann computing paradigm \gls{vpe} micro-architectures with better scheduling have been implemented to overcome memory bottlenecks.
For instance, Vitruvius+ \cite{minervini_vitruvius_2023} tackles out-of-order memory response arrival by improving chaining, allowing intermediate vector blocks to progress ahead of earlier blocks.
A recent effort \cite{saturn} implements improved instruction scheduling mechanisms in Saturn \gls{vpe} for better load balancing of execution pipelines, to address the issue of variable memory latencies, and improve \gls{fpu} utilization. 
Memory architectures tailored to the nature of vector memory accesses can also improve the performance of \glspl{vpe}.
For example, to exploit differences in scalar and vector access patterns, \cite{bicameral} introduces a new cache organization and prefetching mechanisms to reduce memory access latencies. Another work \cite{earth} addresses the inefficiencies in current \glspl{vpe} by implementing a shift-based network in the memory interconnect to speed up constant-strided memory accesses.

However, while these optimizations improve memory latency tolerance and performance of memory operations, insufficient bandwidth imposes a fundamental upper bound on achievable \gls{fpu} utilization. For example, with a 1:1 bandwidth-to-compute ratio, DOTP is fundamentally limited to 50\% of its maximum performance, and AXPY to 33\%.

Beyond traditional architectures, in-memory and near-memory architectures for \glspl{vpe} are witnessing significant interest today because of their ability to exploit the regular nature of vector computations. 
In this emerging computing paradigm, works such as \cite{arcane, csram, mve} add computational units to SRAM memories, thereby reducing the need for expensive data movements to the scalar core. By exploiting the width of cache lines or SRAM rows, these solutions provide benefits of \gls{simd} closer to memory while maintaining a coprocessor abstraction to the scalar core.

Although these efforts help overcome memory access inefficiencies in \glspl{vpe}, they rely on custom core-memory integration rather than focusing on core-side micro-architectural optimizations, which are more generic and readily applicable to current and future \glspl{vpe} designs.

With TROOP, we propose a set of micro-architectural optimizations that allow \glspl{vpe} to exploit a higher memory-bandwidth-to-compute ratio, enabling them to reach their theoretical peak performance on the memory roofline and efficiently minimizing \gls{fpu} underutilization when computing memory-intensive kernels. Our solution is generic and low-cost, as demonstrated by its implementation on a compact RVV-based open-source vector processor.

\section{Background}
\label{sec:backg}

To demonstrate and evaluate TROOP, we implement it on the open-source \gls{rvv}-based Spatz \gls{vpe}, part of the dual-core Spatz cluster \cite{cavalcante2023spatz}. In the following, we describe the architecture of this baseline cluster and Spatz \gls{vpe}, as shown in Figure~\ref{fig: arch}.

\subsection{The Spatz cluster}
The Spatz dual-core cluster consists of two \glspl{cc}, each comprising a Snitch scalar core tightly coupled with a Spatz \gls{vpe}.
Each single-stage RV32I Snitch \cite{Zaruba2020} executes the integer instructions and offloads the scalar and vector floating-point operations to the streamlined, \gls{rvv}-based Spatz \gls{vpe}.
The \glspl{cc} are connected to a 128-KiB \gls{tcdm} through a dedicated XBAR. 

\subsection{Spatz VPE's architecture}
The Spatz architecture is designed to maximize energy efficiency by reducing control logic overhead, directing the energy toward actual computation rather than control operations.

The Controller of the Spatz \gls{vpe} receives the offloaded instructions from Snitch and features \glspl{csr} to track the architectural state (\texttt{vl}, \texttt{vtype}) of the \gls{vpe}.  Within the Controller, the \gls{fpu} sequencer hosts the scalar \gls{fpr} register file and handles the scalar load-store operations, dispatching the remaining scalar floating-point arithmetic and vector operations to the execution units: \gls{vfu}, \gls{vlsu}, and \gls{sldu}. Different execution units can process vector instructions simultaneously while their dependencies are tracked by a scoreboard in the Controller, which orchestrates the vector chaining between dependent instructions by regulating the data movement between the \gls{vrf} and the execution units.

The \gls{vfu} features $F$ \glspl{fpu}, each operating on a 64-bit datapath for an overall computational throughput of $64F$ bits/cycle. This translates to $F$ 64-bit elements/cycle, while narrower data elements are processed in parallel on the 64-bit datapaths in \gls{simd} fashion.
The \glspl{fpu} have a latency of 2 cycles when operating on 32 and 64-bit elements and zero-cycle latency on lower precisions.

Spatz's private \gls{vlsu} performs vector load-store operations through $F$ independent 64-bit interfaces to the L1 \gls{tcdm}, totaling a bandwidth of $64F$ bits/cycle, which matches the computational throughput of the \gls{fpu}. Lastly, the \gls{sldu} facilitates slide and move operations between vector registers.

\begin{figure}
    \centering
    \includegraphics[width=0.9\linewidth]{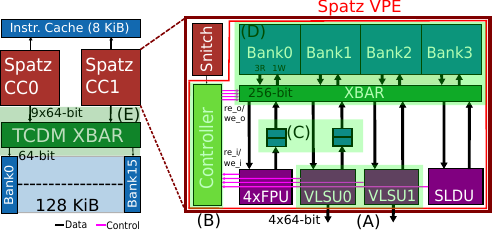}
    \caption{Block diagram of the Spatz cluster architecture, composed of 2 Spatz \glspl{cc}. Each \gls{cc} contains a Snitch and Spatz \gls{vpe}. The \gls{vpe} sub-blocks and the TROOP micro-architectural optimizations (A) to (E) are also shown.}
    \label{fig: arch}
\end{figure}

\subsection{Cluster's Memory Hierarchy}
The cluster's L1 \gls{tcdm} is a \gls{sram} memory consisting of 16 memory banks of size 8KiB with a total of 128KiB shared amongst 2 \gls{cc}. Each bank is 64-bit wide. There is a logarithmic \gls{tcdm} XBAR which provides a single-cycle latency interconnecting $F+1$ ports from each Spatz (5 by default) to the 16 banks.

In addition, each Spatz features a latch-based \gls{vrf}, which acts as an L0 memory within each \gls{vpe}.
The \gls{vrf} consists of 32 vector registers spread across 4 banks. For the default configuration of 4 \glspl{fpu} and a \gls{vlen} of 512 bits, each register occupies 2 consecutive banks. Each bank is $64F$-bit wide and has 3 read ports and 1 write port, all independent. 
Within the \gls{vrf}, an arbiter assigns priorities to each execution unit in case of conflicts on the \gls{vrf} ports, with the \gls{vfu} having the highest priority over the \gls{vlsu} and \gls{sldu}.
The \texttt{vtype} \gls{csr} in the Controller tracks the LMUL parameter, which can assume values in [1,2,4,8] and determines how the \gls{vrf} is logically interpreted. With LMUL=2, for example, the VRF is considered as a 16-register register file, with each register having twice its original size. 
\subsection{Controller and vector chaining}
Vector chaining enables pipelined execution of dependent instructions executed in different functional units. 
In Spatz, the Controller implements chaining by handling data hazards at a VRF-port granularity with a credit-based system. The progress of each instruction is tracked on an element basis instead of across the whole vector length. For simplicity, this is implemented using only a 1-bit credit counter per instruction to track whether the instruction that should execute first has written at least one element in the \gls{vrf} in the current cycle.
When this happens, the Controller allows the dependent instruction to access the \gls{vrf} in the following cycle, consuming the earned credit.
Vector chaining is especially useful between arithmetic (issued to the \gls{vfu}) and memory operations (issued to the \gls{vlsu}) since dependent instructions issued to the same execution unit cannot be executed simultaneously. Also, the current implementation of Spatz's vector chaining works under the hypothesis of a 1:1 memory-bandwidth-to-compute ratio. With TROOP's optimizations, we will streamline the chaining logic to maximize the \gls{vfu} utilization with a 2:1 ratio.

\section{Troop Implementation}
\label{sec:arch}

In this Section, we describe TROOP micro-architectural optimizations in detail, developing them on the Spatz \gls{vpe}.
All of the discussed optimizations-- including decoupled \gls{vlsu} interfaces, improved chaining, shadow buffers, and L0 and L1 memory layouts-- along with their corresponding subblock, are illustrated in Figure \ref{fig: arch}.

\subsection{Decoupled \gls{vlsu} interfaces}
\label{sub: decoupled vlsu}
\glspl{vpe} often features limited bandwidth to memory, penalizing memory-intensive kernels' performance. This happens even if the available memory bandwidth is potentially higher.

For example, in the original Spatz cluster architecture \cite{cavalcante2023spatz} (referred to as Spatz\textsubscript{BASELINE}), the use of $F{=}4$ \gls{vlsu} ports limits the effective L1 memory bandwidth to $8{\times}64$-bit/cycle, whereas the peak bandwidth supported by the L1 \gls{tcdm} is $16{\times}64$-bit/cycle.
This imposes a sub-optimal limit on the achievable \gls{fpu} utilization for common memory-intensive kernels.
To overcome this bottleneck, TROOP features an additional interface in the \gls{vlsu}, effectively doubling the \gls{vlsu} bandwidth to match the peak L1 memory bandwidth.
Although the architecture can theoretically achieve L1 memory roofline with two \gls{vlsu} interfaces (\gls{vlsu}0 and \gls{vlsu}1), the imbalance in the core \gls{vpe} operation limits the achieved performance on memory-intensive kernels.

For example, in Spatz\textsubscript{BASELINE}, both the \gls{vfu} throughput and the \gls{vlsu} bandwidth to L1 operate at $64F$ bits/cycle, maintaining a balanced bandwidth-to-compute ratio.
This balance enables a more straightforward and efficient interleaved access pattern to the \gls{vrf} banks, allowing each execution unit to proceed at maximum throughput.
For instance, a \texttt{vle} operation can write one $64F$-bit word/cycle to $bank_{i+1}\ mod\ 4$, while a simultaneously executing \texttt{vfadd} instruction can write operands to $bank_i\ mod\ 4$. This synchronized, interleaved bank accesses ensure conflict-free execution of the two instructions.

However, with twice the L1 memory bandwidth, the \gls{vlsu} generates two $64F$-bit words/cycle, writing to two consecutive \gls{vrf} banks. In contrast, the \gls{vfu} accesses only a single bank per cycle for reads and writes. 
This mismatch in \gls{vrf} access strides arising from the imbalance in the memory bandwidth and the compute throughput introduces structural conflicts at the \gls{vrf}.
Although increasing the number of banks could alleviate these conflicts, it incurs additional area and energy overheads, which are undesirable in resource-constrained edge devices.

\begin{figure}
    \centering
    \includegraphics[width=0.9\linewidth]{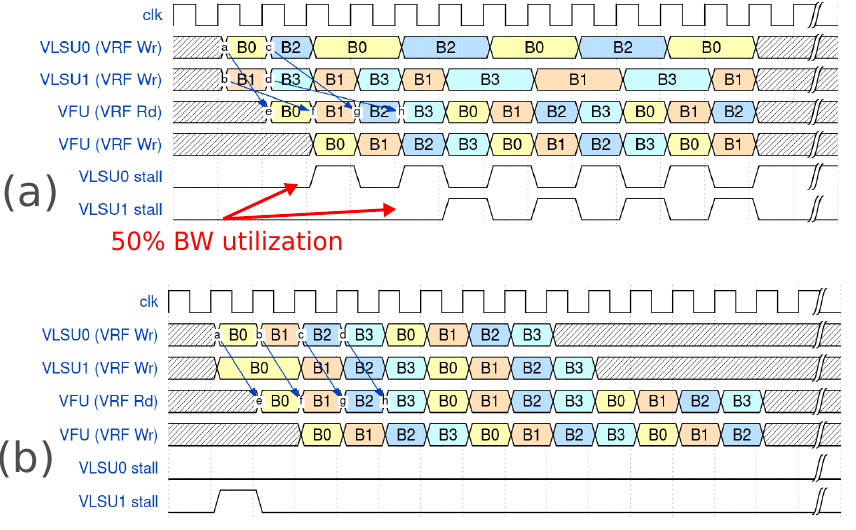}
    \caption{Cycle accurate accesses to the \gls{vrf} banks depicting VLSU stalls (a) without and (b) with decoupling of the interfaces.}
    \label{fig: decouple}
\end{figure}

To address this challenge, we decouple the two \gls{vlsu} interfaces in each Spatz by performing a coarse-grained split of vector memory operations. Both interfaces execute the same load/store instruction, but each operates on a contiguous half of the vector: \gls{vlsu}0 handles the first half, while \gls{vlsu}1 processes the second.
This decoupling enables the \gls{vfu} and both \gls{vlsu} interfaces to access the \gls{vrf} at a stride of one bank per cycle.
This regularizes the \gls{vrf} accesses, allowing the execution units to operate synchronously without conflicts, achieving maximum throughput. 
Figure \ref{fig: decouple} (a) shows a naive bandwidth increase, which results in repeated \gls{vrf} conflicts. In contrast, Figure \ref{fig: decouple} (b) shows how TROOP's \gls{vlsu} decoupling optimization enables conflict-free execution and maximum \gls{fpu} throughput.

\subsection{Improved vector chaining}
\label{sub: chaining}
\begin{figure}
    \centering
    \includegraphics[width=0.9\linewidth]{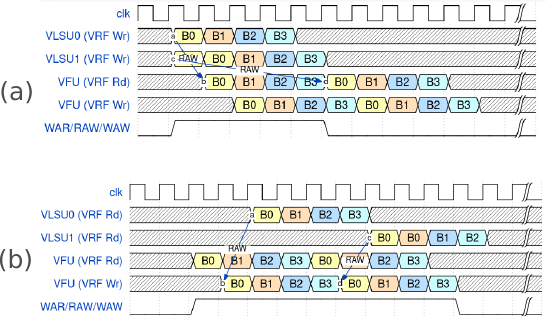}
    \caption{Cycle accurate accesses to \gls{vrf} depicting two vector chaining scenarios between \gls{vlsu} and the \gls{vfu} with different throughput rates.}
    \label{fig: chaining}
\end{figure}

Vector chaining is often handled on an element-wise basis without tracking the overall completion status of vector instructions, as in Spatz\textsubscript{BASELINE}.
While this approach suffices for a balanced bandwidth-to-compute ratio, the introduction of decoupled interfaces necessitates an improved chaining mechanism.
To address this, we add a completion counter to track the progress of vector instruction execution. This allows slower execution units to chain to the desired interface of the \gls{vlsu} once half of the vector operations are completed.
Figure \ref{fig: chaining} depicts the enhanced vector chaining mechanism, showing two scenarios: 
(a) fast$\rightarrow$slow, where a slower \gls{vfu} instruction is chained to a faster \gls{vlsu} instruction and (b) slow$\rightarrow$fast, where a faster \gls{vlsu} instruction chains to a slower \gls{vfu} instruction.

\subsection{Handling \gls{vrf} conflicts}
\label{sub: buffers}
The following optimization targets the higher number of conflicts on the \gls{vrf} write ports caused by the additional \gls{vlsu} interface. In the original Spatz architecture, the \gls{vrf} banks have 3 read and 1 write port, avoiding conflicts between reads and writes.
However, write requests to the same \gls{vrf} bank can conflict. In such cases, the \gls{vrf} prioritizes the \gls{vfu} over the \gls{vlsu} and \gls{sldu}.
While this static priority scheme is sufficient to achieve ideal \gls{fpu} utilization for compute-intensive workloads, it performs poorly for memory-intensive workloads, as discussed below.

\begin{figure}
    \centering
    \includegraphics[width=0.9\linewidth]{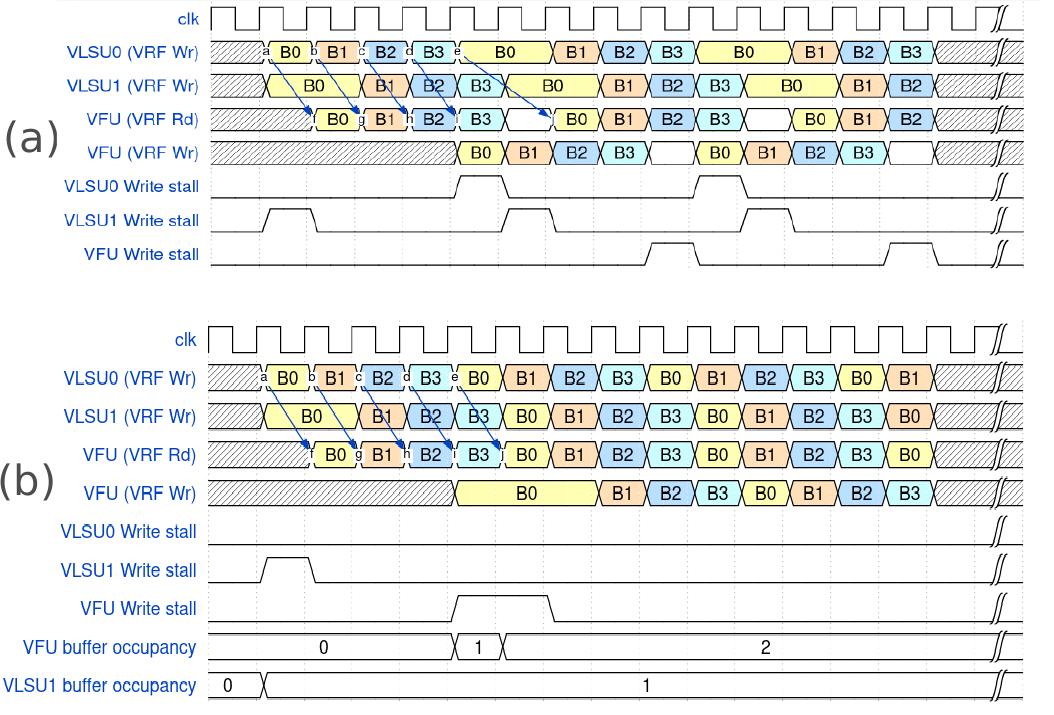}
    \caption{Cycle accurate accesses to the VRF depicting the FPU and VLSU stalls comparing (a) static priority assignment and (b) dynamic priority optimizations with shadow buffers.}
    \label{fig: priority}
\end{figure}

Consider the scenario where a \texttt{vfmacc} operation issues \gls{vrf} reads that are chained to the \gls{vrf} writes from a \texttt{vle} executing in the \gls{vlsu}. 
The controller's chaining logic forwards every \gls{vfu} read request to the \gls{vrf} as long as a write by the \gls{vlsu} has completed. 
However, concurrent writes by the \gls{vfu} and the \gls{vlsu} can conflict, impacting performance.
For example, assuming an \gls{fpu} datapath latency of 2 cycles and an additional 1-cycle write-back pipeline stage, the \gls{vfu} incurs a \gls{vrf} read-to-write latency of 3 cycles. 
Adding a 1-cycle delay because of chaining results in writes from the \gls{vfu} and the \gls{vlsu} to the same \gls{vrf} bank in the same cycle. 
And since the \gls{vfu} is prioritized, its write request proceeds while the \gls{vlsu} writes are stalled.
Because the \gls{vfu} read operand waits for the \gls{vlsu} write, this introduces a 1-cycle bubble in the \gls{vfu} pipeline, reducing \gls{fpu} utilization to 80\%, as shown in Figure \ref{fig: priority} (a).

This exposes a structural limitation in the \gls{vpe} micro-architecture, where the latencies of execution units match the number of \gls{vrf} banks, exacerbating timing conflicts.
Although this issue also exists in Spatz\textsubscript{BASELINE}, it is negligible for compute-intensive kernels where the \gls{vfu} operations dominate the execution time. In contrast, memory-intensive kernels expose this limitation, as peak bandwidth becomes the primary bottleneck.

To overcome this latency-induced conflict, TROOP introduces two optimizations.

\begin{itemize}
    \item \textbf{Dynamic priority allocation} of the \gls{vlsu} interfaces over the \gls{vfu} for writes to the \gls{vrf}. This lets \gls{vlsu} writes complete without stalls, allowing dependent \gls{vfu} reads to proceed in the following cycle.
    After a defined period, priority is restored to the \gls{vfu} to avoid indefinite starvation.
    \item \textbf{Shadow buffers} that temporarily store \gls{vfu} write data to prevent stalling \gls{vfu} writes while prioritizing the \gls{vlsu}. These buffers hide write conflicts and reduce backpressure on the \gls{vfu}, allowing uninterrupted execution of subsequent micro-operations.
\end{itemize}

The dynamic priority allocation and shadow buffer optimizations avoid \gls{vfu} stalling even if the results are still not committed to the \gls{vrf}, maximizing the out-of-order execution capabilities of vector processors.
The \gls{vlsu} is prioritized over the \gls{vfu} as long as the \gls{vfu} writes can be buffered. Since the worst case scenario happens when the \gls{vfu} writes collide with the two \gls{vlsu} interfaces in consecutive cycles, a small buffer of size 2 is sufficient.
This buffer is placed in the \gls{vfu}-to-\gls{vrf} path before the controller, which acts as a centralized point for dependency checks for all \gls{vrf} accesses, knowing when a write is actually committed into the \gls{vrf}.
Additionally, due to the decoupled nature of the \gls{vlsu} interfaces, \gls{vlsu}0 and \gls{vlsu}1 writes can also conflict during the first \gls{vrf} write of a memory instruction. TROOP resolves this using the same principle and introducing a similar buffer in the \gls{vlsu}1-to-\gls{vrf} write path.
Figure \ref{fig: priority} (b) shows the timing diagram of the \gls{vrf} accesses after these optimizations, demonstrating conflict-free writes for all units after the initial conflict phase.

\subsection{L0 \gls{vrf} memory layout} 
\label{sub: L0}
The mapping of vector elements in consecutive logical addresses to the physical banks of the \gls{vrf} is crucial to fully leverage the bandwidth of the \gls{vpe}'s \gls{vlsu}. 
The original Spatz architecture implements a barber-pole \gls{vrf} layout where every set of eight contiguous architectural registers maps the vector elements starting from the same bank, and each consecutive set adds a 1-bank offset to this starting point. 
However, our evaluations show that this layout provides no performance benefits. This is primarily because each bank can support three simultaneous read operands required for a \texttt{vfmacc} operation, therefore not requiring multi-bank accesses.
Barber-pole layout can potentially avoid the initial write conflicts with the \gls{vlsu} for an \gls{fpu} latency of 2, as described in \ref{sub: buffers}.
However, leveraging this requires the programmer to know the \gls{fpu} latency and carefully select the registers in the vectorized code, which introduces non-trivial programming effort.

Instead, our implementation adopts a standard \gls{vrf} layout in which every register is aligned to $bank_0$ or $bank_2$. We resolve the initial conflicts arising from this layout through dynamic priority scheduling and low-cost buffering-- hardware optimizations that are fully transparent to the software.

\subsection{Memory interconnect address scrambling}
Achieving maximum \gls{vrf} bandwidth utilization through the previously described micro-architectural optimizations is useful only if coupled with a full utilization of the memory bandwidth.
This is especially important with multiple decoupled \gls{vlsu} interfaces, when the chances of accessing the same \gls{tcdm} bank are higher.
To prevent the L1 memory from becoming the bottleneck, we introduce address scrambling in the L1 \gls{tcdm} interconnect to achieve maximum bandwidth utilization for different \gls{lmul} configurations supported by \gls{rvv}.
For instance, in a \gls{vrf} configuration with \gls{lmul}=1 or 2, each vector register occupies 512- or 1024-bits, respectively. These fit within a row of the L1 memory spanning 16 banks (each 64 bits wide), meaning that the accesses from the two \gls{vlsu} interfaces cannot conflict when accessing the \gls{tcdm}. 
However, for \gls{lmul}=4 or 8, accesses from the interfaces map to the same banks, causing conflicts.

To overcome this, we add an address scrambling scheme in the \gls{tcdm} interconnect to offset by 8 banks every second and third row in a group of 4 rows in the \gls{tcdm}. 
This misalignment prevents the two \gls{vlsu} interfaces from targeting the same \gls{tcdm} banks, ensuring conflict-free access and maximizing L1 bandwidth utilization across all \gls{lmul} settings.
While it may be argued that the \gls{tcdm} bank conflicts naturally resolve after the initial cycle, this non-intrusive scrambling scheme allows for a simple decoupled \gls{vlsu} implementation that maximizes performance without the need of independent \gls{vlsu} units executing different instructions in parallel.

\subsection{Software optimizations}
With the improved vector chaining shown in Figure \ref{fig: chaining}, instructions following the fast$\rightarrow$slow dependency can achieve maximum throughput. However, for slow$\rightarrow$fast instructions, the progress of the faster instruction is limited by the slower instruction.
To mitigate this, we propose loop unrolling and software pipelining techniques to expose more parallelism and achieve the maximum throughput for the faster \gls{vlsu} operations.
For example, during AXPY, the vector store operation \texttt{vse} following the \texttt{vfmacc} instruction cannot utilize the available bandwidth of both the \gls{vlsu} interfaces.
With a loop unrolling factor of 2, we break this dependency between store and compute operations, allowing the \gls{vlsu} to proceed without dependency operating at full throughput.

\subsection{Reduction optimizations}
To fully leverage the higher memory bandwidth during a DOTP, the \gls{vpe}'s reduction logic should be optimized and able to sustain full throughput. This is not always the case, as in Spatz\textsubscript{BASELINE}, since, originally, the main bottleneck was the lower memory bandwidth.
However, reduction cycles heavily impact the execution time in the enhanced Spatz architecture with doubled memory bandwidth.
Therefore, we optimize the reduction operator and implement it in $log2$ number of reduction steps, further improving \gls{fpu} utilization during reductions.

\section{Evaluation}
\label{sec:eval}

\subsection{Evaluation Setup}
We implement TROOP micro-architectural optimizations in SystemVerilog and integrate them into the energy-efficient, streamlined open-source Spatz \gls{vpe}. We benchmark TROOP on key kernels targeting linear algebra, signal processing, and machine learning, including lower bit-precisions for embedded applications.
We consider three different configurations of the Spatz cluster for our evaluations:
\begin{itemize}
    \item Spatz\textsubscript{BASELINE}: Default configuration with $64F$-bit/cycle bandwidth to \gls{tcdm}.
    \item Spatz\textsubscript{2$\times$BW}: Spatz with double bandwidth to the \gls{tcdm} ($128F$-bit/cycle) without any micro-architectural optimizations.
    \item Spatz\textsubscript{2$\times$BW-TROOP}: Spatz with double the bandwidth to \gls{tcdm} as above, but enhanced with TROOP optimizations.
\end{itemize}
We use QuestaSim-64 2021.3\_2 for cycle-accurate simulations and report \gls{fpu} utilization (\% of execution cycles where \gls{fpu} is in operation). 

To evaluate TROOP's impact on the \gls{ppa} metrics, we perform the physical implementation of all three configurations in a 12nm technology node and compare their \gls{ppa} metrics.
For a fair \gls{ppa} comparison, our Spatz\textsubscript{BASELINE} also includes reduction optimizations. 

\begin{figure}
    \centering
    \includegraphics[width=1\linewidth]{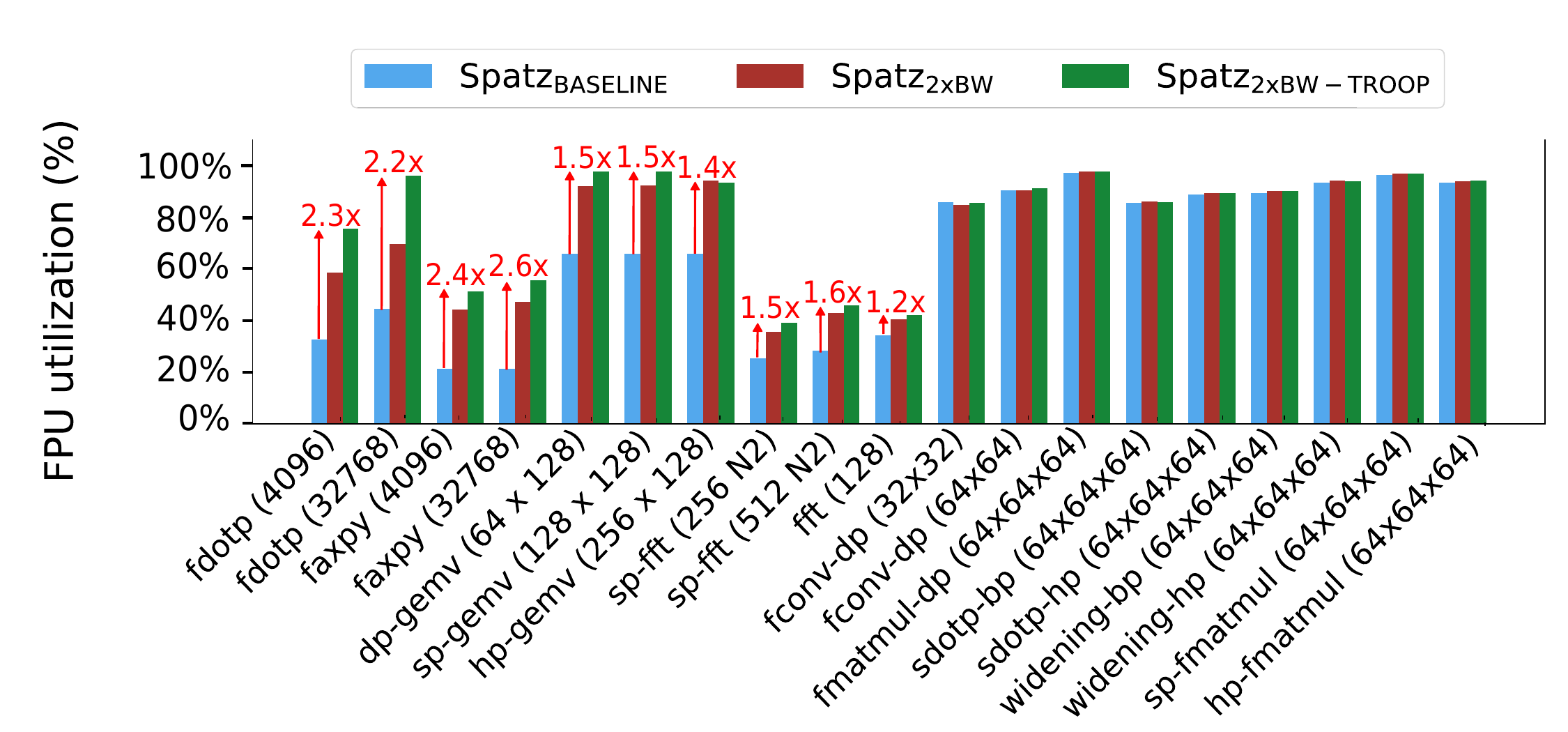}
    \caption{FPU utilization for different kernels and Spatz cluster configurations.}
    \label{fig: perf}
\end{figure}

\subsection{Performance Comparison}
Figure \ref{fig: perf} shows the achieved FPU utilization for the three Spatz cluster configurations when executing our kernels.

\subsubsection{Spatz\textsubscript{2$\times$BW}}
The performance improvement of Spatz\textsubscript{2$\times$BW} over the baseline is primarily due to the increased bandwidth to the shared L1 \gls{tcdm}. Memory-intensive kernels such as DOTP and AXPY achieve a speedup of 1.8$\times$ and 2.1$\times$ over the baseline, respectively, for an application vector length of 4096 elements. Similarly, GEMV and FFT also achieve speedups of up to 1.5$\times$.
As expected, compute-intensive kernels' performance remains constant as they are not limited by memory bandwidth.

While we can improve the performance of memory-intensive kernels by doubling the bandwidth to L1, the achieved \gls{fpu} utilizations are well below the theoretical roofline peak. For instance, GEMV reaches an \gls{fpu} utilization of only 92\% despite being compute-bound on Spatz\textsubscript{2$\times $BW}. 
The situation is worse for DOTP and AXPY, whose \gls{fpu} utilizations peak at 59\% and 44\%, far from the theoretical maximum of 100\% and 66\%, respectively.

These results highlight a fundamental micro-architectural limitation in the current Spatz \gls{vpe} design: while compute-bound workloads can approach peak performance, memory-bound workloads cannot fully leverage the increased bandwidth to the shared L1 \gls{tcdm}, leaving the \gls{fpu} underutilized.

\subsubsection{Spatz\textsubscript{2$\times$BW-TROOP}}
We evaluate the performance impact of TROOP on Spatz with twice the bandwidth to the shared L1 \gls{tcdm}. As shown in Figure \ref{fig: perf}, DOTP achieves an \gls{fpu} utilization of 76\%, a significant improvement from 59\% achieved by Spatz\textsubscript{2$\times$BW}. This is because TROOP regularizes the \gls{vrf} accesses, eliminating structural repetitive conflicts. At longer vector lengths, we can also amortize the reduction costs and achieve close to ideal utilization of 96\% for DOTP, whereas the Spatz\textsubscript{2$\times$BW} is limited to 70\% utilization.

TROOP is transparent to the datapath latencies of the \gls{vfu}. For example, double- and single-precision GEMV kernels achieve ideal \gls{fpu} utilization. TROOP's optimized priority allocation and shadow buffers allow hiding the write conflicts at the \gls{vrf} arising from an \gls{fpu} latency of 2 cycles for single and double-precision floating-point operations.
In contrast, conflicts do not occur for half-precision operations with no latency, so TROOP has no visible effect on performance.

For AXPY, a combination of TROOP and loop unrolling pushes the \gls{fpu} utilization to 55\%, i.e., 2.6$\times$ higher than the baseline. However, ideal \gls{fpu} utilizations can only be achieved with a bandwidth-to-compute ratio of at least 3:1.

\subsection{Physical Implementation}

\begin{figure}
    \centering
    \includegraphics[width=0.8\linewidth]{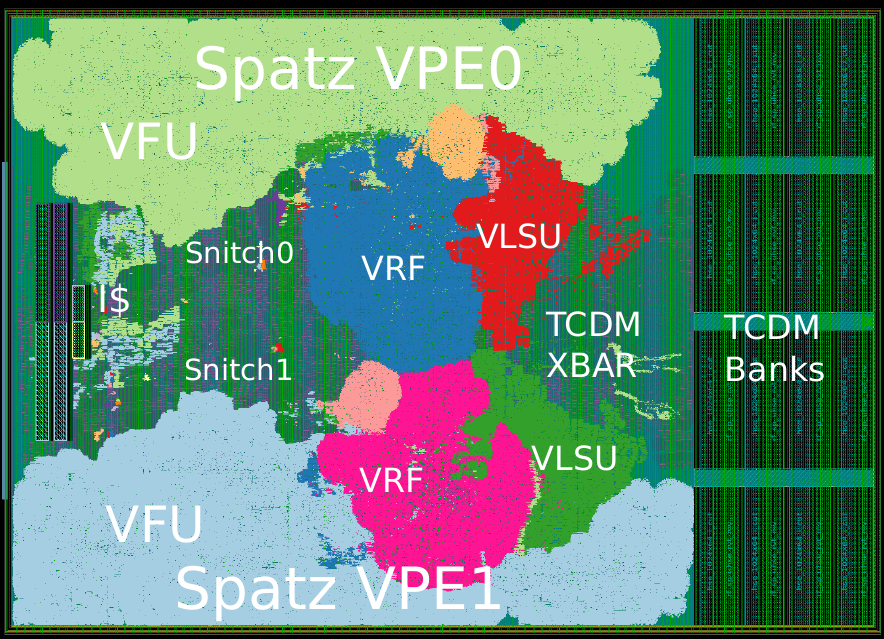}
    \caption{Floorplan of the Spatz cluster implementation in gf12nm showing various sub-blocks of the \gls{vpe} (\gls{vfu}, \gls{vrf}, \gls{vlsu}) along with the TCDM Banks and XBAR Interconnect.}
    \label{fig: fp}
\end{figure}

We perform the physical implementation of Spatz\textsubscript{BASELINE} and Spatz\textsubscript{2$\times$BW-TROOP} on a 12nm technology node using Fusion Compiler 2022.03sp2. We synthesize and PnR (Place and Route) the design targeting a frequency of 1 GHz in the slow corner (SS/125C/0.72V). 
The final floorplan of Spatz\textsubscript{2$\times$BW-TROOP} after PnR is shown in Figure \ref{fig: fp}.

\subsubsection{Area Comparison}
Table \ref{tab:spatz_area_comparison} reports the area results.
Overall, the area impact of the TROOP set of micro-architectural optimizations on the cluster is limited to 7\%. 

At the cluster level, the \gls{tcdm} interconnect increases by 1.8$\times$ due to the larger crossbar connecting 18 Spatz ports instead of 10 to the 16 \gls{tcdm} banks.
The remaining optimizations, such as coarse decoupling within the \gls{vlsu}, address scrambling in the \gls{tcdm} interconnect, and \gls{vrf} with standard layout instead of barber pole, only introduce minor modifications to the address generation control logic and do not impact the overall area.

At the \gls{vpe} level, the Spatz \gls{vlsu} area increases by a factor of 2.6$\times$ compared to the Spatz\textsubscript{BASELINE}. This is due to the doubling of the number of interfaces to the L1 \gls{tcdm} and the L0 \gls{vrf}, contributing to most of the area increase for the \gls{vpe}.
The three additional ports to the \gls{vrf} (1RD+1WR for data and 1RD for indices) from the \gls{vlsu} increase the \gls{vrf} area by 4\%. 
The controller area, which now also hosts shadow buffers to buffer \gls{vfu} and \gls{vlsu}1 writes during \gls{vrf} conflicts, sees an increase of 40 kGE. However, the overall contribution of the controller is only 1.2\% of the total area of the Spatz cluster.

\subsubsection{Timing Comparison}

TROOP does not introduce any new critical paths, ensuring Spatz\textsubscript{2$\times$BW- TROOP} achieves a maximum frequency of 1.23 GHz in the typical corner (TT/0.8V/25C), the same achieved by the original Spatz\textsubscript{BASELINE} architecture.

\begin{table}
    \centering
    \caption{Area breakdown of evaluated Spatz configurations}
    \begin{tabular}{rrr}
         \toprule
          Cell area [kGE] & Spatz\textsubscript{BASELINE} & Spatz\textsubscript{2$\times$BW-TROOP} \\
          (*$\times$) &  &  \\
          \midrule
         \multicolumn{3}{l}{\textbf{Spatz cluster}} \\
         iCache & 150 (1.00$\times$) & 149 (1.00$\times$) \\
         TCDM Banks & 1190 (1.00$\times$) & 1190 (1.00$\times$) \\
         TCDM XBAR & 53 (1.00$\times$) & 94 (1.78$\times$) \\
         \midrule
         \multicolumn{3}{l}{\textbf{Spatz CC $\times$ 2}} \\
         Snitch & 48 (1.00$\times$) &  47 (1.00$\times$) \\
         VRF & 403 (1.00$\times$) & 420 (1.04$\times$) \\
         VFU & 1610 (1.00$\times$) & 1592 (1.00$\times$) \\
         VLSU & 115 (1.00$\times$) & 298 (2.58$\times$) \\
         Controller & 12 (1.00$\times$) & 53 (4.46$\times$) \\
         Other & 514 (1.00$\times$) & 533 (1.04$\times$) \\
         \midrule
         \textbf{TOTAL} & 4095 (1.00$\times$) & 4377 (1.07$\times$) \\
         \bottomrule
    \multicolumn{3}{l}{* Scaling factor normalized to Spatz\textsubscript{BASELINE}}\\
    \end{tabular}

    \label{tab:spatz_area_comparison}
\end{table}

\subsection{Energy Efficiency Comparison}

Table \ref{tab:spatz_eneff_comparison} reports the energy efficiency of the evaluated Spatz cluster configurations at similar area utilizations ($\sim$60\%). 
We run the above evaluated benchmarks on a back-annotated netlist after place-and-route and generate the \gls{vcd} for the kernel execution period. 
We generate the source data for our kernels using a normal distribution with a mean of 0 and a variance of 1. 
The generated \gls{vcd} along with the parasitics and constraints are provided to the PrimeTime-2022.03 compiler, and power estimations are done in the typical corner (TT/0.8V/25C) at 1 GHz.

Spatz\textsubscript{2$\times$BW-TROOP} significantly improves the energy efficiency of key low operational intensity kernels such as AXPY and DOTP by 26\% and 45\%, respectively.
GEMV also sees an increase in efficiency by 9\% achieving 52 DP-GFLOPs/W.
Importantly, these gains are achieved without compromising the energy efficiency for compute-intensive kernels such as convolutions and \gls{gemm}, thanks to the low-cost nature of TROOP optimizations.

\begin{table}
    \centering
    \caption{Energy efficiency comparison of Spatz cluster configurations at 1 GHz (TT/0.8/25C)}
    \begin{tabular}{rrrr}
         \toprule
         Energy Efficiency & Spatz\textsubscript{BASELINE} & Spatz\textsubscript{2$\times$BW-TROOP} \\
          (GFLOPs/W) &  &  \\
          \midrule
         dp-faxpy & 21.8 & 27.5 (\textbf{1.26x}) \\
         dp-fdotp & 25.9 & 37.5 (\textbf{1.45x}) \\
         dp-gemv & 48.0 & 51.8 (\textbf{1.09x}) \\
         sp-gemv & 98.9 & 107.2 (\textbf{1.08x}) \\
         dp-fft & 23.6 & 24.0 (\textbf{1.02x}) \\
         sp-fft & 44.6 & 45.7 (\textbf{1.02x}) \\
         dp-fconv2d & 60.9 & 61.0 (1.00x) \\
         dp-fmatmul & 61.1 & 61.1 (1.00x) \\
         sp-fmatmul & 125.8 & 124.9 (1.00x) \\
         \bottomrule
    \end{tabular}

    \label{tab:spatz_eneff_comparison}
\end{table}

\subsection{Roofline comparison with SoA}

\begin{figure}
    \centering
    \includegraphics[width=1.0\linewidth]{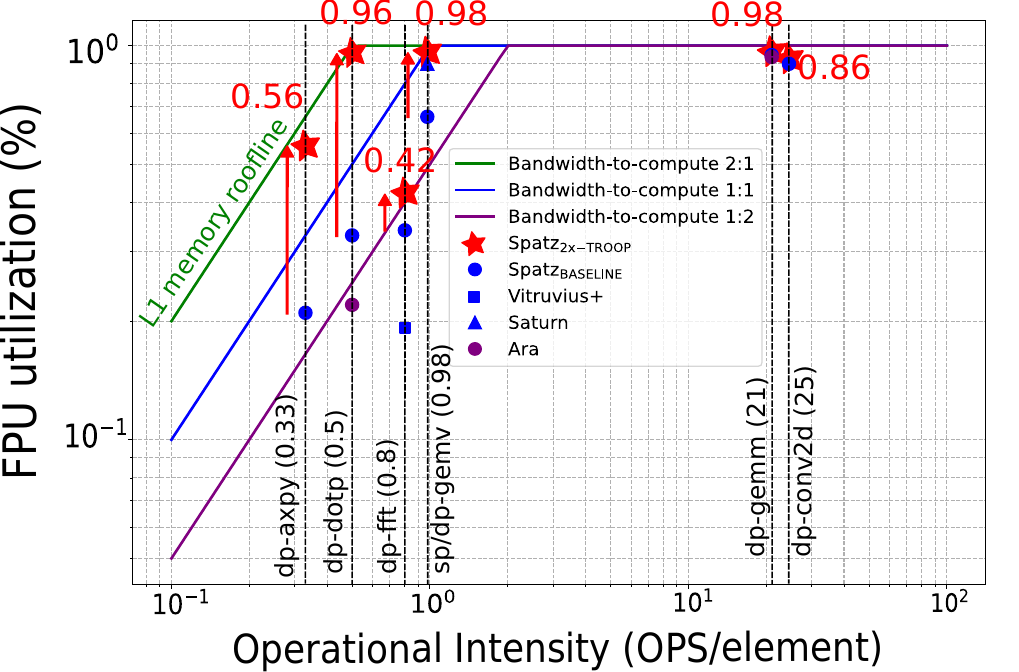}
    \caption{Normalized roofline-plot comparison of achieved FPU utilization for state-of-the-art \glspl{vpe} for different kernels.}
    \label{fig: roofline}
\end{figure}

State-of-the-art \glspl{vpe} achieve lower \gls{fpu} utilizations for memory-intensive workloads due to the low bandwidth-to-compute ratios.
\todo{For instance, Saturn reports a utilization of 90\% for the crucial GEMV kernel.}
The baseline Spatz can only achieve an \gls{fpu} utilization of 33\% and 21\% for the DOTP and AXPY kernels. Vitruvius+, with similar bandwidth-to-compute ratios, reports a peak performance utilization of 38\% for AXPY.
As an example of a \gls{vpe} with even lower bandwidth-to-compute ratio, Ara achieves only a utilization of 22\% for DOTP.
\todo{Although optimizations done in Saturn and Vitruvius+ improve memory-latency tolerance, the primary bottleneck for memory-intensive kernels remains the limited memory bandwidth of the \glspl{vpe}, which we increase in this work.}

Figure \ref{fig: roofline} presents a normalized roofline comparison for different state-of-the-art \glspl{vpe} and different kernels.
The y-axis reports the \gls{fpu} utilization, while the x-axis plots the operational intensity (OPS per element), enabling normalized comparison across different architectures and kernel precisions. For consistency, the bandwidth-to-compute ratio is used rather than the absolute bandwidth.
Compared to the state-of-the-art, Spatz\textsubscript{2$\times$BW-TROOP} features twice the bandwidth-to-compute ratio to accelerate memory-intensive workloads. Additionally, Spatz\textsubscript{2$\times$BW-TROOP} achieves utilizations of 96\%, 55\%, and 98\% for DOTP, AXPY, and GEMV, respectively, demonstrating that TROOP is a valid aid to reach theoretical L1 memory roofline performance for key memory-intensive kernels.

\section{Conclusions}
\label{sec:conclusions}

State-of-the-art \glspl{vpe} achieve close-to-ideal \gls{fpu} utilization on compute-intensive kernels but fall short on kernels with low operational intensity commonly used in \gls{ai} due to their limited bandwidth with respect to their compute capabilities.
In this work, we present TROOP, a set of low-cost micro-architectural optimizations to fully leverage the higher available bandwidth of shared L1-memory clusters and achieve peak \gls{fpu} utilization on memory-intensive kernels.
We demonstrate TROOP on the compact state-of-the-art Spatz \gls{vpe}, with significant speedups up to 2.6$\times$ (AXPY), reaching at-the-roofline performance for key memory-intensive kernels like DOTP, AXPY, and GEMV.
TROOP also boosts the \gls{vpe}'s energy efficiency, e.g., by up to 45\% for DOTP (38 DP-GFLOPs/W), without compromising the energy efficiency of compute-intensive kernels such as \gls{gemm} and convolution (61 DP-GFLOPs/W), incurring a small area overhead of 7\%.

\ifreviewmode
\else
\section{Acknowledgments}
\label{sec:ack}
\changes{This work has received funding from the Swiss State Secretariat for Education, Research, and Innovation (SERI) under the SwissChips initiative.}

\fi 

\bibliographystyle{IEEEtran}
\bibliography{references,ieeetran}

\begin{thebibliography}{10}
\providecommand{\url}[1]{#1}
\csname url@samestyle\endcsname
\providecommand{\newblock}{\relax}
\providecommand{\bibinfo}[2]{#2}
\providecommand{\BIBentrySTDinterwordspacing}{\spaceskip=0pt\relax}
\providecommand{\BIBentryALTinterwordstretchfactor}{4}
\providecommand{\BIBentryALTinterwordspacing}{\spaceskip=\fontdimen2\font plus
\BIBentryALTinterwordstretchfactor\fontdimen3\font minus \fontdimen4\font\relax}
\providecommand{\BIBforeignlanguage}[2]{{%
\expandafter\ifx\csname l@#1\endcsname\relax
\typeout{** WARNING: IEEEtran.bst: No hyphenation pattern has been}%
\typeout{** loaded for the language `#1'. Using the pattern for}%
\typeout{** the default language instead.}%
\else
\language=\csname l@#1\endcsname
\fi
#2}}
\providecommand{\BIBdecl}{\relax}
\BIBdecl

\bibitem{ai_memorywall}
A.~Gholami \emph{et~al.}, ``{AI and Memory Wall},'' \emph{IEEE Micro}, vol.~44, no.~3, pp. 33--39, 2024.

\bibitem{yuan2025native}
\BIBentryALTinterwordspacing
J.~Yuan \emph{et~al.}, ``{Native Sparse Attention: Hardware-Aligned and Natively Trainable Sparse Attention},'' 2025. [Online]. Available: \url{https://arxiv.org/abs/2502.11089}
\BIBentrySTDinterwordspacing

\bibitem{speed_dnn}
C.~Wang \emph{et~al.}, ``{SPEED: A Scalable RISC-V Vector Processor Enabling Efficient Multiprecision DNN Inference},'' \emph{IEEE Transactions on Very Large Scale Integration (VLSI) Systems}, vol.~33, no.~1, pp. 207--220, 2025.

\bibitem{eyerissv2}
\BIBentryALTinterwordspacing
Y.-H. Chen, T.-J. Yang, J.~Emer, and V.~Sze, ``{Eyeriss v2: A Flexible Accelerator for Emerging Deep Neural Networks on Mobile Devices},'' 2019. [Online]. Available: \url{https://arxiv.org/abs/1807.07928}
\BIBentrySTDinterwordspacing

\bibitem{flexnn}
\BIBentryALTinterwordspacing
A.~Raha, D.~A. Mathaikutty, S.~K. Ghosh, and S.~Kundu, ``{FlexNN: A Dataflow-aware Flexible Deep Learning Accelerator for Energy-Efficient Edge Devices},'' 2024. [Online]. Available: \url{https://arxiv.org/abs/2403.09026}
\BIBentrySTDinterwordspacing

\bibitem{tsmc3d}
\BIBentryALTinterwordspacing
``{{3DFabric}},'' {TSMC}, accessed May 15, 2025. [Online]. Available: \url{https://3dfabric.tsmc.com/english/dedicatedFoundry/technology/3DFabric.htm}
\BIBentrySTDinterwordspacing

\bibitem{cmos2.0}
\BIBentryALTinterwordspacing
``{{The CMOS 2.0 Revolution}},'' {IMEC}, accessed May 15, 2025. [Online]. Available: \url{https://www.imec-int.com/en/articles/cmos-20-revolution}
\BIBentrySTDinterwordspacing

\bibitem{allinone}
\BIBentryALTinterwordspacing
``{{All-in-one RISC-V AI IP}},'' {Semidynamics.}, accessed April 04, 2025. [Online]. Available: \url{https://semidynamics.com/en/file/pb/j2hwj4nan1n52tzy15pltjzp63vpaj}
\BIBentrySTDinterwordspacing

\bibitem{tensix}
\BIBentryALTinterwordspacing
``{{TT Architecture and Metalium Guide}},'' {Tenstorrent}, accessed May 09, 2025. [Online]. Available: \url{https://github.com/tenstorrent/tt-metal/blob/main/METALIUM_GUIDE.md}
\BIBentrySTDinterwordspacing

\bibitem{mi300x}
A.~Smith and V.~Alla, ``{AMD Instinct MI300X Generative AI Accelerator and Platform Architecture},'' in \emph{2024 IEEE Hot Chips 36 Symposium (HCS)}, 2024, pp. 1--22.

\bibitem{cavalcante2023spatz}
M.~Perotti, S.~Riedel, M.~Cavalcante, and L.~Benini, ``Spatz: Clustering compact risc-v-based vector units to maximize computing efficiency,'' \emph{IEEE Transactions on Computer-Aided Design of Integrated Circuits and Systems}, vol.~44, no.~7, pp. 2488--2502, 2025.

\bibitem{saturn}
\BIBentryALTinterwordspacing
J.~Zhao \emph{et~al.}, ``{Instruction Scheduling in the Saturn Vector Unit},'' 2024. [Online]. Available: \url{https://arxiv.org/abs/2412.00997}
\BIBentrySTDinterwordspacing

\bibitem{minervini_vitruvius_2023}
F.~Minervini \emph{et~al.}, ``Vitruvius+: {An} {Area}-{Efficient} {RISC}-{V} {Decoupled} {Vector} {Coprocessor} for {High} {Performance} {Computing} {Applications},'' \emph{ACM Transactions on Architecture and Code Optimization}, vol.~20, no.~2, pp. 1--25, 2023.

\bibitem{ara2perotti24}
M.~Perotti \emph{et~al.}, ``Ara2: Exploring single- and multi-core vector processing with an efficient {RVV} 1.0 compliant open-source processor,'' \emph{IEEE Transactions on Computers}, vol.~73, no.~7, pp. 1822--1836, 2024.

\bibitem{Takahashi_2023}
K.~Takahashi \emph{et~al.}, ``{Performance Evaluation of a Next-Generation SX-Aurora TSUBASA Vector Supercomputer},'' in \emph{High Performance Computing: 38th International Conference, ISC High Performance}, 2023, p. 359–378.

\bibitem{bicameral}
S.~Rebolledo, B.~Perez, J.~L. Bosque, and P.~Hsu, ``{The Bicameral Cache: a split cache for vector architectures},'' in \emph{2024 IEEE 30th International Conference on Parallel and Distributed Systems (ICPADS)}, 2024, pp. 728--736.

\bibitem{earth}
\BIBentryALTinterwordspacing
H.~Guan \emph{et~al.}, ``{Efficient Architecture for RISC-V Vector Memory Access},'' 2025. [Online]. Available: \url{https://arxiv.org/abs/2504.08334}
\BIBentrySTDinterwordspacing

\bibitem{arcane}
\BIBentryALTinterwordspacing
V.~Petrolo \emph{et~al.}, ``{ARCANE: Adaptive RISC-V Cache Architecture for Near-memory Extensions},'' 2025. [Online]. Available: \url{https://arxiv.org/abs/2504.02533}
\BIBentrySTDinterwordspacing

\bibitem{csram}
M.~Kooli \emph{et~al.}, ``{Towards a Truly Integrated Vector Processing Unit for Memory-bound Applications Based on a Cost-competitive Computational SRAM Design Solution},'' \emph{J. Emerg. Technol. Comput. Syst.}, vol.~18, no.~2, Apr. 2022.

\bibitem{mve}
\BIBentryALTinterwordspacing
A.~Khadem \emph{et~al.}, ``{Multi-Dimensional Vector ISA Extension for Mobile In-Cache Computing},'' 2025. [Online]. Available: \url{https://arxiv.org/abs/2501.09902}
\BIBentrySTDinterwordspacing

\bibitem{Zaruba2020}
F.~Zaruba, F.~Schuiki, T.~Hoefler, and L.~Benini, ``Snitch: A tiny pseudo dual-issue processor for area and energy efficient execution of floating-point intensive workloads,'' \emph{IEEE Transactions on Computers}, vol.~70, no.~11, pp. 1845--1860, 2021.

\end{thebibliography}

\end{document}